\documentclass[10pt,conference]{IEEEtran}

\usepackage{cite}
\usepackage{amsmath,amssymb,amsfonts}
\usepackage{algorithmic}
\usepackage{graphicx}
\usepackage{textcomp}

\def\BibTeX{{\rm B\kern-.05em{\sc i\kern-.025em b}\kern-.08em
    T\kern-.1667em\lower.7ex\hbox{E}\kern-.125emX}}
\usepackage{url}    
\usepackage{tikz}
\newcommand*\circled[1]{\tikz[baseline=(char.base)]{
        \node[shape=circle,draw,inner sep=0.5pt] (char) {#1};}}
        
\begin{document}

\title{Enhancing the Clique Local Decoder to Correct Length-2 Space Errors in the Surface Code
}

\author{\IEEEauthorblockN{Zikang Jia}
\IEEEauthorblockA{\textit{Department of Mathematics} \\
\textit{University of Michigan}\\
Ann Arbor, USA \\
zkjia@umich.edu}
\and
\IEEEauthorblockN{Shravan Veerapaneni}
\IEEEauthorblockA{\textit{Department of Mathematics} \\
\textit{University of Michigan}\\
Ann Arbor, USA \\
shravan@umich.edu}
\and
\IEEEauthorblockN{Gokul Subramanian Ravi}
\IEEEauthorblockA{\textit{Computer Science and Engineering} \\
\textit{University of Michigan}\\
Ann Arbor, USA \\
gsravi@umich.edu}
}

\maketitle

\begin{abstract}
The growing demand for fault-tolerant quantum computing drives the need for efficient, scalable Quantum Error Correction (QEC) strategies. Conventional decoders designed for worst-case error scenarios incur significant overhead, prompting the development of local decoders, that leverage the sparse and often trivial nature of many quantum errors, to support the conventional decoders. The previously proposed \texttt{Clique} decoder addresses this by handling isolated, length‑1 space and time errors within the cryogenic environment with minimal hardware costs, thereby mitigating I/O bandwidth constraints between cryogenic quantum systems and room-temperature processors.

Building on this foundation, we propose \texttt{Clique\_L2} that extends the Clique-based approach by relaxing some original constraints and incorporating additional low-cost logic to also correct length‑2 error chains in space, which become non-trivial occurrences at higher physical error rates and code distances. This enhanced capability not only further reduces out-of-the-fridge data transmission but also adapts more effectively to clustered errors observed under a variety of noise models. Specifically, under data-qubit-only errors and uniformly random noise, \texttt{Clique\_L2} achieves up to 8.95x decoding bandwidth reduction over the original  \texttt{Clique} (or   \texttt{Clique\_L1}) decoder, especially beneficial at higher code distances. When clustered errors and longer error chains are more likely to occur, \texttt{Clique\_L2} 
achieves up to 18.3x decoding bandwidth reduction over \texttt{Clique\_L1}, achieving substantial benefits across a wide range of physical qubit error rates.
\end{abstract}


\section{Introduction}
Quantum computing introduces a groundbreaking computational approach, harnessing quantum phenomena like superposition, entanglement, and interference to solve problems beyond classical reach. While algorithms such as Shor’s factoring~\cite{Shor_1997} and Grover’s search~\cite{Grover96afast} hold immense promise, their real-world implementation is hindered by the fragile nature of quantum states. Present-day quantum devices, known as Noisy Intermediate-Scale Quantum (NISQ) systems~\cite{preskill2018quantum}, are prone to errors such as decoherence, gate imperfections, and measurement inaccuracies. Overcoming these challenges requires the implementation of Quantum Error Correction (QEC) techniques to achieve fault-tolerant quantum computation~\cite{chuang1997prescription}.

Quantum Error Correction (QEC) enhances robustness by encoding logical qubits across multiple physical qubits, enabling error correction through repeated measurements and classical decoding. Among the leading QEC techniques, the surface code stands out due to a high error threshold and local connectivity, making it well-suited for practical implementation~\cite{Fowler_2012}. However, a major challenge lies in the classical decoding of error syndromes, which becomes a significant bottleneck, particularly in cryogenic quantum systems such as those utilizing superconducting qubits.  

The conventional approach to QEC decoding involves performing all classical processing at room temperature~\cite{Dennis_2002}, outside the dilution refrigerator. However, this method faces severe I/O bandwidth constraints, as the high volume of error data must be transmitted across the cryogenic barrier within a limited timeframe~\cite{AFS}—an obstacle that conventional cryogenic setups at scale will struggle to address. An alternative approach is to shift the classical processing entirely into the cryogenic domain~\cite{holmes2020nisq+,QECOOL}, thereby mitigating bandwidth limitations. Yet, this solution introduces its own set of challenges, as it is constrained by stringent area, power, and thermal budgets inherent to cryogenic environments.  
These scalability challenges underscore the necessity for more efficient and adaptive classical processing strategies. This work specifically focuses on optimizing the approach to QEC decoding to address these limitations.

Decoders are traditionally designed to handle rare worst-case error scenarios, ensuring quantum execution accuracy but at the expense of significant resource overhead. For example, state-of-the-art minimum weight perfect matching (MWPM) decoders~\cite{Dennis_2002} have explored leveraging multi-core and multi-threaded computing to accelerate the decoding process~\cite{higgott2025sparse}. However, an analysis of error distributions reveals that a large portion of errors occurring during quantum computation are sparse and isolated~\cite{ravi2023better}. These errors are inherently `trivial' to decode and do not necessitate computationally expensive decoding techniques.  
This disparity presents an opportunity to optimize QEC decoding by decoupling the processing of common, trivial errors from rare, complex ones. By adopting a more adaptive approach, the burden on conventional decoding resources can be significantly reduced, improving overall efficiency without compromising fault tolerance.

Building on this philosophy, previous work by Ravi et al.~\cite{ravi2023better} introduced the hardware `Clique' decoder. Clique was tailored for the surface code, and was designed to be implemented with Single-Flux Quantum (SFQ) logic~\cite{ERSFQ}, making it suitable for the cryogenic environment. Clique acts as a first-level decoder and decodes only `length-1' errors leaving all other errors to be handled by a conventional room-temperature decoder. It requires only minimal cryogenic hardware—approximately 10 combinational logic gates per physical qubit—and successfully corrects large fractions of total error signatures. While Clique is always effective in reducing the error bandwidth, its efficacy is reduced when physical error rates and/or the QEC code distance are high. Thus, the ability to also correct higher length errors locally at reasonable cost would be very beneficial.

In this work we identify that by relaxing a few hardware constraints in the original Clique decoder, which we refer to as \texttt{Clique\_L1}, and with the addition of minimal logic, we can enhance a Clique-like decoder to also tackle length-2 errors in space (i.e. when two data errors occur adjust to each other, forming a chain of length 2). We we refer to this enhanced local decoder as \texttt{Clique\_L2}.
Our evaluations with phenomenological noise models show that \texttt{Clique\_L2} considerably improves upon the decoding coverage achieved by \texttt{Clique\_L1} at minimal additional cost. We further evaluate \texttt{Clique\_L2} under specialized noise models that represent scenarios in which space errors are correlated and prone to clustering, and showcase even greater potential of \texttt{Clique\_L2} in such scenarios. Specifically, we use a `Gaussian' model that increases the probability of data qubit errors in the neighborhood of an existing error, and  a `Dual-Error' model that randomly produces errors on pairs of data qubits that share a stabilizer in the surface code. 

To summarize, the \texttt{Clique\_L2} decoder significantly reduces the I/O QEC bandwidth in the following aspects:

\begin{itemize}
    \item Under scenarios involving only data qubit errors (i.e. a code capacity model), \texttt{Clique\_L2} significantly reduces bandwidth usage by handling 97–99\% of decoding tasks locally across a wide range of physical error rates—especially at higher code distances—achieving up to an 8.95x improvement over \texttt{Clique\_L1}.

    \item When dealing with uniformly random noise, \texttt{Clique\_L2} consistently decreases the need to send decoding tasks out of the cryogenic domain, achieving up to a 1.44x improvement over \texttt{Clique\_L1}. Its efficiency increases with code distance, offering meaningful bandwidth savings under general noisy conditions.
    
    \item With densely clustered errors under a Gaussian-style noise model, \texttt{Clique\_L2} dramatically reduces the fraction of decoding tasks sent to room-temperature by up to 8.60x compared to  \texttt{Clique\_L1}, highlighting its effectiveness particularly at higher code distances and complex error scenarios.
    
    \item Under an orchestrated `Dual-Error' noise model, where correlated length-2 error chains are frequent, \texttt{Clique\_L2} outperforms \texttt{Clique\_L1} by handling up to 18.38x more decodes locally, substantially reducing bandwidth requirements across code distances.
    
\end{itemize}

\section{Background}
\label{bkg}
\subsection{QEC Overview}
QEC leverages redundant encoding of quantum information, along with repeated measurements on ancillary qubits, to track and correct physical errors without collapsing the logical state. Each logical qubit is encoded into a block of physical qubits, often called \textbf{data qubits}. Additional ancilla qubits are then entangled with these data qubits to extract information about any errors that may occur, without destroying the underlying quantum state. Specifically, the ancilla qubits are measured to produce classical bits referred to as \textbf{syndromes}, which indicate the presence of particular Pauli errors (bit-flips X, phase-flips Z, both Y, or none I). A \textbf{decoder} then interprets these syndromes to localize and identify the errors, allowing the quantum system to implement corrections. If the underlying physical qubit error rates and gate fidelities are below a certain threshold (depending on the code and decoder), scaling up the code distance (i.e., the size of the block encoding) drives the logical error rate exponentially lower~\cite{9906129, Fowler_2012}. 

\subsection{Surface Codes}
\label{b:surface}

\begin{figure}[t]
\centering
\includegraphics[width=0.65\columnwidth,trim={1.85cm 1.75cm 0.025cm 0.05cm},clip]{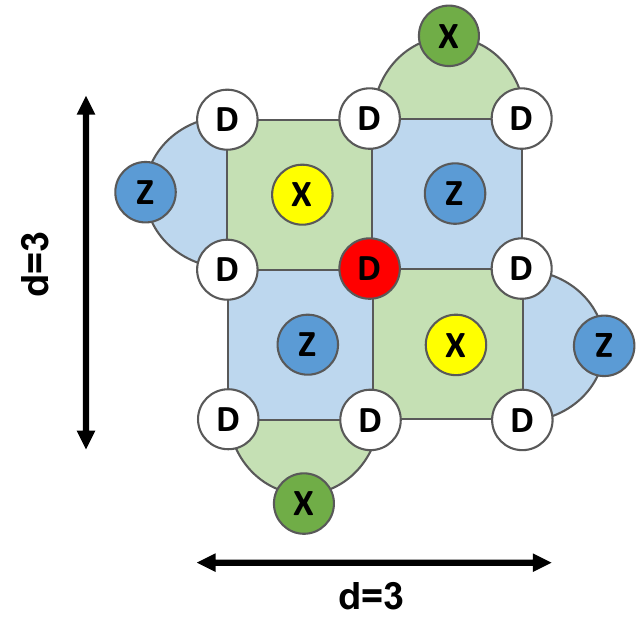}
\caption{
 Distance 3 rotated surface code detecting Z error on central data qubit D by flipping neighboring X ancillae.
}
\label{fig:clique_bkg}
\end{figure}

There are several well-known quantum error correction (QEC) codes that protect quantum information by encoding a single logical qubit into multiple physical qubits. One of the promising approaches is the surface code—a type of topological code that arranges qubits on a two-dimensional grid and uses only local interactions for error detection and correction~\cite{das2022lilliput}. It allows physical qubit error rate to be closer to the 0.1-1\% range~\cite{Fowler_2012}, meaning that if each physical operation can be made accurate 99-99.9\% of the time, logical error rates can be suppressed exponentially by increasing the lattice size. 

More specifically, the two-dimensional surface code lattice consists of alternating physical data and \textbf{parity} (ancilla) qubits. The total number of qubits required to encode a single logical qubit scales quadratically with the code distance, $d$. This code distance sets a fundamental constraint on the length of error chains that can be corrected.

For example, Fig.~\ref{fig:clique_bkg} depicts a surface code layout with code distance $d=3$. Data qubits are shown as white ‘D’ circles, while ancilla qubits for bit-flip and phase-flip error detection are marked as green ‘X’ and blue ‘Z’, respectively. A phase-flip (‘Z’) error on the red-highlighted data qubit is detected by its diagonally adjacent ‘X’ ancillae (shaded in yellow). Similarly, bit-flip (‘X’) errors are identified by neighboring ‘Z’ ancillae, and ‘Y’ errors trigger both. Repeated entanglement of data and ancilla qubits via stabilizer circuits enables syndrome extraction for error detection.

\subsection{QEC Decoding}
A crucial challenge in quantum error correction (QEC) is real-time \textit{decoding}, which processes parity-check data to track and correct errors on logical qubits, ensuring smooth quantum execution.  
Decoding efficiency is critical—if data production outpaces decoding, a \textit{backlog} forms, causing an \emph{exponential} slowdown~\cite{terhal2015quantum}. Optimizing \textit{throughput} (error processing rate) and \textit{latency} (decoding time) is essential, especially for \textit{blocking operations} like non-Clifford gates, where delays increase idle time and the risk of irrecoverable errors.  

Significant progress has been made in QEC decoding. The \textit{minimum-weight perfect matching (MWPM)} decoder, widely used for surface codes, has been optimized from $O(N^3 \log N)$ to $O(N^2)$. The \textit{Union-Find decoder} offers near-linear runtime and hardware implementation but with slightly lower accuracy. Further advances stem from algorithmic innovations~\cite{delfosse2021almost, higgott2025sparse, wu2023fusion} and targeted optimizations~\cite{ravi2023better, vittal2023astrea, alavisamani2024promatch}.  
Yet, real-time decoding at scale remains challenging. A {million-qubit quantum computer} is expected to generate {1 terabit per second} of measurement data, pushing even the best decoders to their limits.

\subsection{Considerations for Cryogenic Quantum Systems}

Quantum devices—particularly those based on superconducting qubits—operate at extremely low temperatures (typically 10–20 millikelvin) to preserve quantum coherence. As a result, the supporting classical electronics must either interface from room temperature or be adapted to function within the cryogenic environment.  

Full-fledged decoding for QEC is both high-performance and power-intensive, making room-temperature placement the conventional choice. However, cryogenic I/O constraints significantly limit this approach. QEC error syndromes are generated approximately every microsecond—or faster, depending on qubit readout capabilities—requiring immediate identification and correction. As code distances grow, the required syndrome transmission bandwidth per logical qubit reaches gigabits per second. In large-scale systems with high code distances, this translates to an overall transmission bandwidth demand in the terabits per second range~\cite{ravi2023better,AFS}. Meeting this demand is highly challenging due to limited I/O wiring availability. While coaxial cables can facilitate data transmission between the dilution refrigerator and room-temperature electronics, cryogenic systems impose strict constraints on space, thermal dissipation, and signal integrity, particularly as qubit counts rise.  

Some prior works have explored fully cryogenic QEC decoding to bypass I/O bottlenecks. For instance, NISQ+~\cite{holmes2020nisq+} performs all decoding within the cryogenic domain, effectively alleviating bandwidth constraints. However, fully cryogenic decoders face severe limitations in chip area, thermal management, and computational complexity, creating another resource bottleneck. Similar challenges are observed in recent cryogenic decoders like QECOOL~\cite{QECOOL}, where resource constraints become even more pronounced as logical qubit counts and code distances increase—key requirements for fault-tolerant quantum computing.  

Both room-temperature and fully cryogenic decoding approaches present fundamental scalability challenges, underscoring the need for innovative alternative solutions.

\section{Lightweight QEC Decoding with \texttt{Clique\_L1}}
\label{motive}

\subsection{Motivation}
\label{BTWC_motive}

In practice, physical errors in QEC occur sparsely across space and time, making long error chains exceptionally rare. Given a physical qubit error rate of $p$, the probability of an error chain spanning $k$ qubits scales as $O(p^{k})$. Quantinuum recently announced a milestone by achieving a two-qubit gate error rate of approximately $8.6 \times 10^{-4}$ on its H1-1 system~\cite{Quantinuum-1}, with future improvements expected to lower this to around $1 \times 10^{-4}$ in 2029~\cite{Quantinuum-2}.  

Ravi et al.~\cite{ravi2023better} highlighted that isolated errors, and more generally, short error chains, can be efficiently decoded using simple localized logic. In contrast, longer chains create complex syndrome patterns across the logical qubit block, necessitating more intensive decoding. This distinction suggests a natural opportunity to decouple the handling of common, simple QEC error signatures from the decoding of rare, complex errors.

To address this challenge at a first order, Ravi et al.~\cite{ravi2023better} introduced the Clique local decoder, referred to as \texttt{Clique\_L1} in this work, as a first-level decoding solution for the surface code. Designed to handle isolated errors (length-1 error chains), \texttt{Clique\_L1} operates in a cryogenic environment (4K) using Single-Flux-Quantum (SFQ) logic~\cite{ERSFQ}. It requires minimal hardware—approximately 10 combinational logic gates per physical qubit—and successfully corrects 70–99\% of total error signatures, depending on physical error rates, code distances, and other factors.  
Beyond correcting simple errors, \texttt{Clique\_L1} also identifies complex errors (length-2 or longer chains) and offloads them to a more powerful room-temperature decoder for further processing. This filtering mechanism significantly reduces the error data transmitted across the cryogenic barrier, alleviating I/O bandwidth bottlenecks while preserving worst-case decoding capabilities.  
Next, we delve deeper into the \texttt{Clique\_L1} design, as our work builds upon its foundation.

\subsection{Design of \texttt{Clique\_L1}}
\label{Clique1_design}

\begin{figure}[t]
\centering
\includegraphics[width=0.95\columnwidth,trim={0cm 0cm 0cm 0cm},clip]{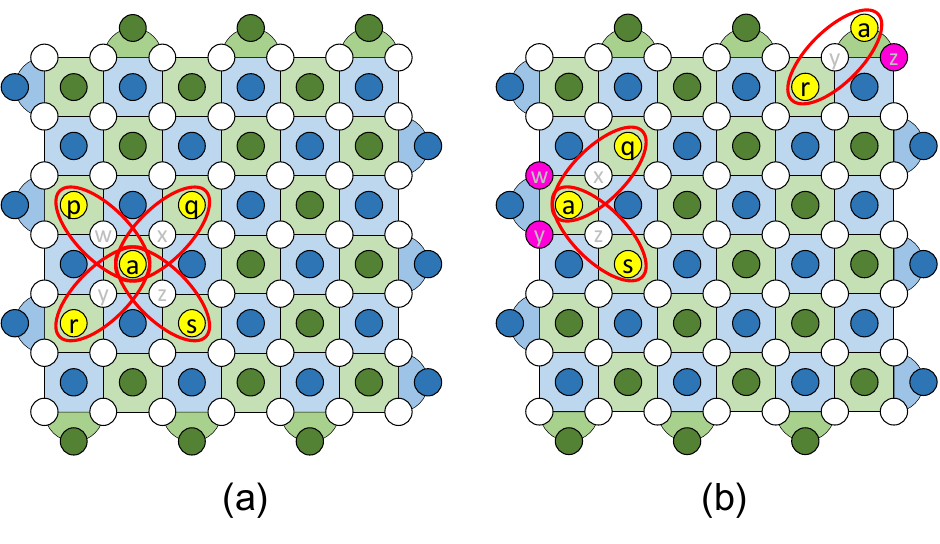}
\caption{
\texttt{Clique\_L1} evaluates the syndromes within each local clique region and determines whether the error pattern is simple enough to be decoded locally. If yes, the corresponding correction can be applied to the neighboring data qubit.}
\label{fig:old_Clique_design}
\end{figure}

The functionality of the \texttt{Clique\_L1} decoder is realized by analyzing the parity syndromes from each local `Clique' of ancilla qubits. Fig.~\ref{fig:old_Clique_design} presents examples of a surface-code-based logical qubit with $d=7$. The `X'-type ancillae are labeled in green, while the `Z'-type ancillae are labeled in blue. In our example, we focus on the `X'-type ancillae, as the same logic applies to the `Z'-type due to the nature of CSS codes. 

As shown in the figure, when a central ancilla qubit detects an error, it becomes an `active' clique. For every active clique, the parity of its immediate surrounding ancilla qubits is relevant to the local decoding process. For instance, if the ancilla qubit `a' in Fig.~\ref{fig:old_Clique_design}(a) is set, the \texttt{Clique\_L1} decoder checks the parity of its neighboring qubits `p', `q', `r', and `s' (at most four). 

The decoder categorizes the error signature based on the parity of these neighboring qubits. \textbf{Even parity:} If none or two of the neighbors are set, the error signature is classified as \emph{complex}. In this case, the \texttt{Clique\_L1} decoder lacks sufficient local information to resolve the error with high confidence, necessitating the use of a full decoder. \textbf{Odd parity:} If one or three neighbors are set, the error signature can be locally decoded and corrected using \texttt{Clique\_L1}. For example, if the active clique at `a' reports a single neighboring set syndrome at `q', the likely error is on the data qubit `x' in the middle, which should be corrected. In such cases, the local syndrome information is sufficient for high-confidence decoding and correction.

The \texttt{Clique\_L1} logic varies when handling edge and corner cases. In these scenarios, even when the neighborhood parity is even, the decoder can sometimes perform local corrections. For instance, in the top-right corner of the surface code in Fig.~\ref{fig:old_Clique_design}(b), for the active clique centered at `a', if the neighborhood parity is even (i.e., `r' is unset), \texttt{Clique\_L1} can still locally decode the error to be on data qubit `z' (labeled in pink), without requiring a full decoder. Further, on the left edge of Fig.~\ref{fig:old_Clique_design}(b), if neither `q' nor `s' are set (even neighborhood parity), the local decoder will correct either `w' or `y', as both corrections are equivalent up to a stabilizer.

\texttt{Clique\_L1} can effectively manage measurement errors by analyzing results from multiple measurement rounds. Measurement errors may cause qubits to flip randomly, but to distinguish actual data errors from transient measurement errors, parity qubits are measured for multiple rounds. Errors that persist across these rounds are classified as data errors, while those that disappear are dismissed as temporary measurement errors. This approach is similar to the idea of increasing the spatial code distance in error correction—performing more rounds enhances the reliability of identifying true data errors. \texttt{Clique\_L1} typically employs two measurement rounds to balance accuracy and hardware costs. However, additional rounds can be used if necessary for more confidence in error detection.

Therefore, the design of \texttt{Clique\_L1} enables it to efficiently handle a variety of error scenarios while minimizing reliance on global decoding mechanisms.


\subsection{Limitations of \texttt{Clique\_L1}}

Although a large fraction of syndromes can be handled trivially—especially at very low physical error rates and relatively small code distances—\texttt{Clique\_L1} is designed specifically to decode and correct isolated data or measurement errors, referred to as length-1 errors.
However, the effectiveness of Clique can especially decrease at higher error rates and code distances, and especially in noise scenarios  characterized by errors that are correlated and tend to cluster. In such scenarios, erroneous qubits are more likely to be surrounded by additional errors—\texttt{Clique\_L1} will be unable to decode these. 
Such scenarios are realistic and can occur due to the physical syndrome check circuit constructions, chip defects~\cite{fuhui2023codesign,smith2022scalingsuperconductingquantumcomputers}, etc.
In such cases, a larger fraction of syndromes must be transmitted to the full-fledged decoder for resolution.

In this work, we demonstrate that with minor logic modifications, \texttt{Clique\_L1}'s capability can be extended to decode error chains in space of length 2 (measurement error chains are still length 1), where two adjacent data qubits experience errors. While generally achieving broader error coverage than \texttt{Clique\_L1}, our enhanced decoder, termed \texttt{Clique\_L2}, is particularly beneficial in environments where errors cluster.

\section{\texttt{Clique\_L2}: Length-2 Decoding of Errors}

\subsection{Intuition and Logic for Length-2 Decoding}
\label{logic_l2}
When attempting to correct errors through local decoding, we have previously seen that \texttt{Clique\_L1} is capable of handling length-1 errors, i.e., $k=1$. With a few simple modifications, the correction capabilities of this local decoder can be extended to support length-2 errors specifically in space, leading to our \texttt{Clique\_L2} design.

The key modifications in \texttt{Clique\_L2} involve changes to the `activation' condition of the clique and adjustments to the neighborhood syndrome parity constraints. Recall that in \texttt{Clique\_L1}, local decoding was only performed when the central syndrome was set, and an odd neighborhood syndrome parity was required to execute the correction. 
In the proposed \texttt{Clique\_L2} design, we relax the \textbf{activation constraint}, allowing decoding even when the central syndrome is \textit{unset} (i.e., the clique is `inactive'). However, in such cases, local decoding is performed under the \textbf{opposite} neighborhood syndrome parity condition compared to \texttt{Clique\_L1}. Specifically, when the central syndrome is unset, local decoding occurs in the case of an \textbf{even} neighborhood syndrome parity, effectively enabling the correction of length-2 error chains in space.

\begin{figure}[t]
\centering
\includegraphics[width=0.95\columnwidth,trim={0cm 0cm 0cm 0cm},clip]{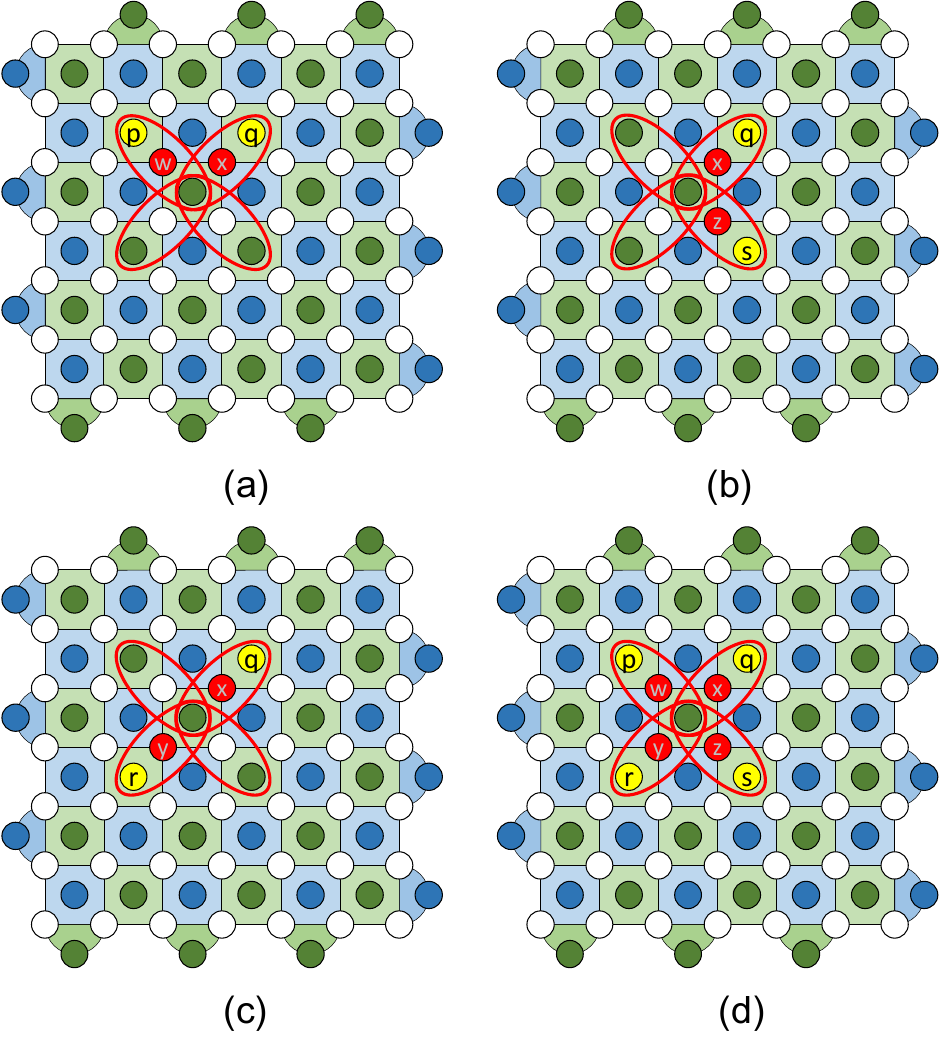}
\caption{
 Scenarios with space error lengths $k \geq$ 2 that are detected and corrected with \texttt{Clique\_L2}: (a) horizontal, (b) vertical, (c) diagonal, (d) mixed. 
}
\label{fig:new_clique_intuition}
\end{figure}

To illustrate the intuition behind the proposed design, we present multiple scenarios in Fig.~\ref{fig:new_clique_intuition}. The figure depicts four typical length-2 space error cases that can be locally decoded and corrected using \texttt{Clique\_L2}: horizontal, vertical, diagonal, and mixed cases.

In the case (a) horizontal, \texttt{Clique\_L2} detects that two parity qubits 'p' and 'q' are set (labeled in yellow) in the same row and the centered qubit is unset. Then the likely error is at the horizontal data qubits 'w' and 'x' labeled in red, which should be corrected. In the case (b) vertical, \texttt{Clique\_L2} detects that two parity qubits 'q' and 's' are set (labeled in yellow) in the same column and the centered qubit is unset. Then the horizontal data qubits 'x' and 'z' labeled in red need to be corrected. Besides, \texttt{Clique\_L2} can detect syndromes in the diagonal scenarios (case (c)). For example, if two qubits 'q' and 'r' are set along the diagonal and detected, the diagonal data qubits 'x' and 'y' should be corrected. The mixed scenario (case (d)) will rarely be present, where all parity qubits 'p', 'q', 'r', and 's' reachable by \texttt{Clique\_L2} are set. In this situation, all neighboring data qubits 'w', 'x', 'y', and 'z' need to be corrected. Note that the other three cases will occur much more frequently, and cover all scenarios for length-2 space errors. 

\subsection{Designing the \texttt{Clique\_L2} Decoder}

To enable length-2 space error decoding with the \texttt{Clique\_L2} decoder, certain modifications to the \texttt{Clique\_L1} logic are required. However, several challenges must be addressed to prevent erroneous decoding. The key challenges are as follows:

\begin{enumerate}
    \item Since length-1 errors must still be decoded alongside length-2 errors, the overall local decoding needs to support both cases. This presents a challenge because, in our Clique-based design, the two cases require conflicting conditions for local decoding: 
    \begin{enumerate}
        \item Length-1 decoding typically requires the central syndrome in the clique to be set and the neighborhood syndrome parity to be \textbf{odd}. 
        \item Length-2 decoding typically requires the central syndrome to be \textbf{unset} and the neighborhood syndrome parity to be \textbf{even}.
    \end{enumerate}

    \item The \texttt{Clique\_L1} decoding logic classifies length-2 errors as complex error signatures that require full decoding. However, \texttt{Clique\_L2} can process these errors (specifically the ones in space) locally. Thus, the length-1 classification logic must be revised to ensure smooth and synergistic execution.

    \item The special decoding and correction rules for edges and corners in the \texttt{Clique\_L1} decoder must be carefully adapted to avoid conflicts with the enhanced \texttt{Clique\_L2} strategy.
\end{enumerate}

\begin{figure}[t]
\centering
\includegraphics[width=0.95\columnwidth,trim={0cm 0cm 0cm 0cm},clip]{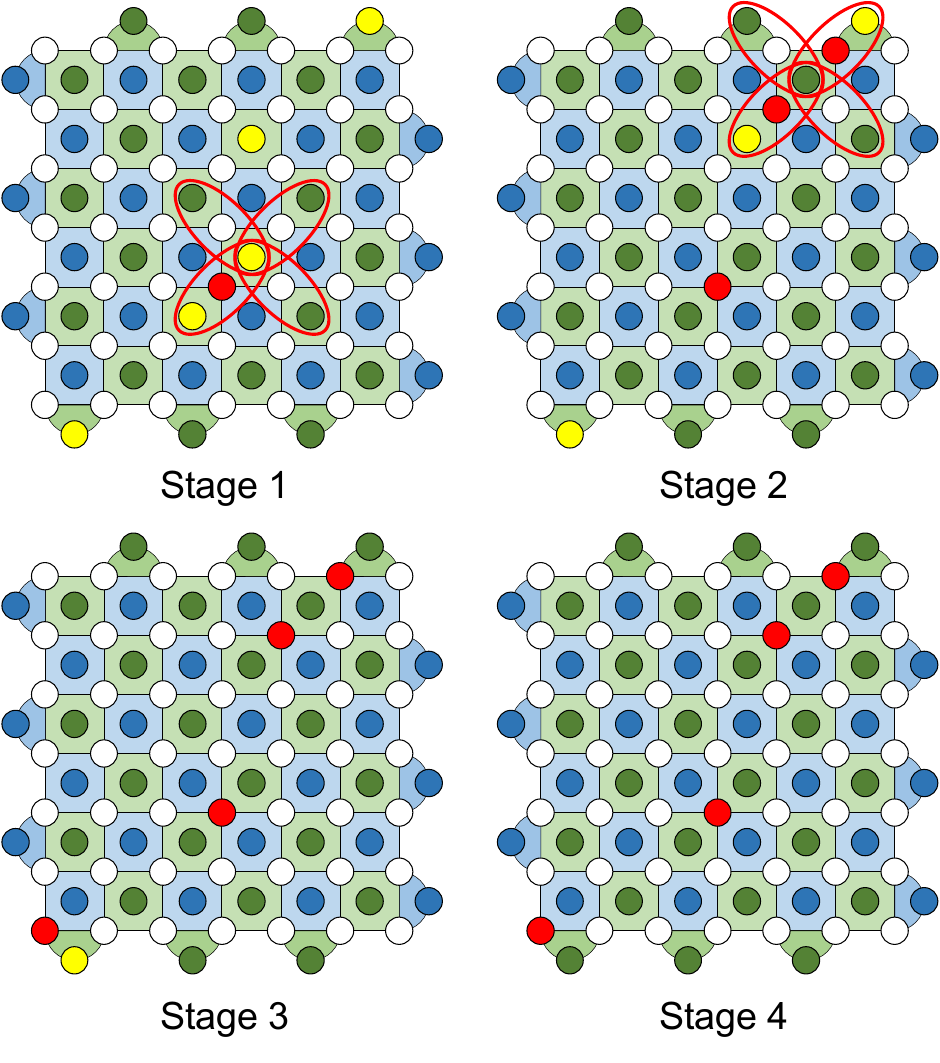}
\caption{
 An example of the sequential four-stage pipeline for \texttt{Clique\_L2}. \\
 Stage 1: Only length-1 error decoding logic of \texttt{Clique\_L1} is activated to decode and correct error chains of length 1, while excluding edge and corner cases. During this stage, syndrome information is updated; however, the algorithm does not classify any scenarios as complex. \\
 Stage 2: With the updated syndrome information on the surface code, only length-2 error decoding logic of \texttt{Clique\_L2} is activated to decode and correct error chains of length 2. \\
 Stage 3: Edge and corner cases are handled by applying the correction logic inherited from \texttt{Clique\_L1} that is tailored for these boundary conditions. \\
 Stage 4: Complex error signatures, which require full-scale decoding at room temperature, are identified. In this situation, the algorithm recognizes the scenario as non-complex and terminates the decoding process. \\
Note that the decoding sequence must be followed in order and cannot be rearranged. For instance, length-2 error decoding must be performed before handling edge and corner cases, i.e., Stage 2 must precede Stage 3. Otherwise, as illustrated in the figure, the top-right and bottom-left corner syndromes would be corrected prematurely, preventing the resolution of the syndrome in the third row that should have been addressed during length-2 decoding.
}
\label{fig:error_correction_k2}
\end{figure}

By addressing and analyzing each challenge, we propose a four-stage sequence to execute the \texttt{Clique\_L2} logic (illustrated with the example in Fig.~\ref{fig:error_correction_k2}):

\begin{enumerate}
    \item \textbf{Stage 1:} Decode and correct error chains of length 1 using the  \texttt{Clique\_L1} decoding logic, excluding edge and corner cases. Update the syndrome information but do not classify scenarios as complex.
    
    \item \textbf{Stage 2:} Decode and correct error chains of length 2 (with logic described in Sec.\ref{logic_l2}) based on the updated syndrome information.
    
    \item \textbf{Stage 3:} Handle all edge and corner cases by applying the corresponding edge- and corner-specific correction logic inherited from \texttt{Clique\_L1}.
    
    \item \textbf{Stage 4:} Identify complex error signatures that require full-fledged decoding at room temperature.
\end{enumerate}

Fig.~\ref{fig:error_correction_k2} illustrates an example syndrome in which five parity-check qubits are activated, posing a nontrivial decoding scenario for existing decoders. With the proposed four-stage sequential decoding pipeline, the syndrome can be progressively decoded and partially corrected by addressing simpler cases, such as length-1 and length-2 errors. It is noteworthy that each stage's partial correction is optimal, as the Clique decoder identifies the shortest possible data error chains. The sequential partial decoding shown in Fig.~\ref{fig:error_correction_k2} eventually achieves complete correction. If any syndrome remains after these stages, it would indicate a complex scenario, which requires further decoding by full-fledged decoders at room temperature.

\subsection{More Decoding Examples}
\label{decoding examples}

\begin{figure}[t]
\centering
\includegraphics[width=0.95\columnwidth,trim={0cm 0cm 0cm 0cm},clip]{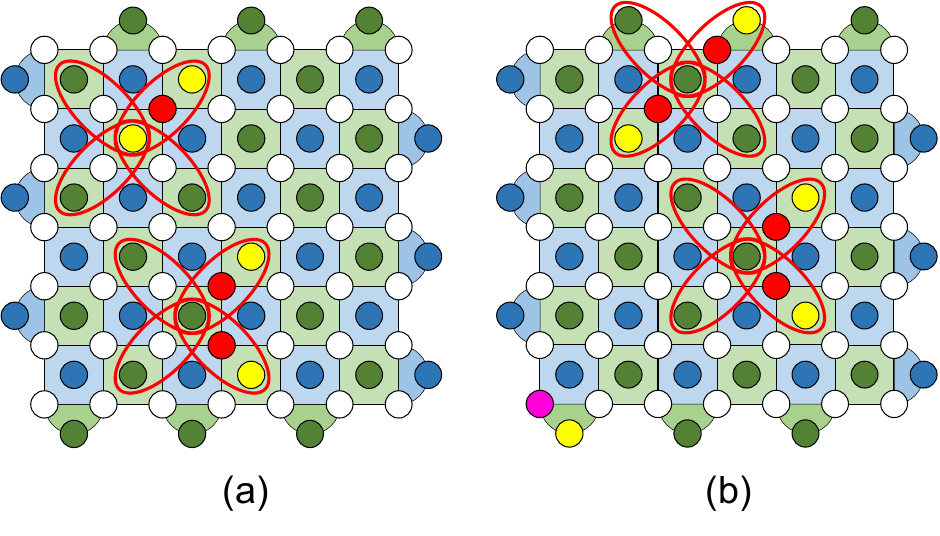}
\caption{Decoding scenarios for \texttt{Clique\_L2} with non-intersecting Cliques: (a) error chains of length 1 and length 2; (b) error chains of length 2 and corner error.
}
\label{fig:decoding_examples}
\end{figure}

Compared to \texttt{Clique\_L1}, \texttt{Clique\_L2} can decode and correct more complex error scenarios locally within the cryogenic domain. In addition to length-1 errors and errors at corners and edges, \texttt{Clique\_L2} handles length-2 space errors and combinations involving length-1, length-2 space, edge, and corner errors. For instance, Fig.~\ref{fig:decoding_examples}(a) illustrates a scenario involving both length-1 and length-2 errors on the surface code, which are sequentially corrected in Stages 1 and 2 of the proposed decoding pipeline. Fig.~\ref{fig:decoding_examples}(b) shows a scenario with two length-2 errors and one corner error, which are decoded and corrected sequentially in Stages 2 and 3 of the pipeline.

\subsection{Decoding Intersecting Cliques}

\begin{figure}[t]
\centering
\includegraphics[width=0.95\columnwidth,trim={0cm 0cm 0cm 0cm},clip]{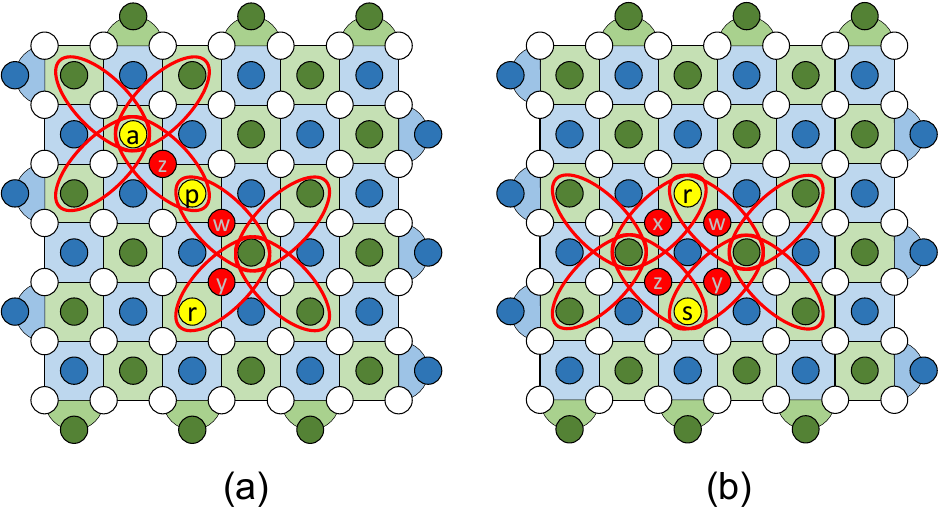}
\caption{Examples illustrating problems of intersecting Cliques: (a) Two Cliques intersect at the parity qubit `p'. The top-left Clique performs length-1 error decoding to flip the data qubit `z', while the bottom-right Clique performs length-2 error decoding to flip data qubits `w' and `y'. (b) Two Cliques intersect at parity qubits `r' and `s'. Here, the left Clique applies length-2 error decoding to flip data qubits `x' and `z', while the right Clique also employs length-2 error decoding to flip data qubits `w' and `y'.
}
\label{fig:intersecting_cliques}
\end{figure}

The examples discussed in Fig.~\ref{fig:decoding_examples} involve decoding and correction using non-intersecting Cliques, meaning that each parity qubit is covered by exactly one Clique. However, problems arise if a parity qubit is included in two Cliques simultaneously. For instance, Fig.~\ref{fig:intersecting_cliques} illustrates scenarios where two Cliques intersect at a single parity qubit, each independently performing its decoding logic.

In scenario (a), two Cliques intersect at one parity qubit `p'. Here, the top-left Clique applies length-1 error decoding, flipping data qubit `z', while the bottom-right Clique applies length-2 error decoding, flipping data qubits `w' and `y'. As a result, the two Cliques together flip a total of three data qubits, leading to an incorrect correction. Such scenarios should be recognized as complex and delegated to a full-fledged decoder.

Scenario (b) involves two Cliques intersecting at two parity qubits along the horizontal direction. In this case, both Cliques independently perform length-2 error decoding, resulting in the data qubits `r' and `s' being flipped. However, this situation should be handled by only one Clique to ensure correct error correction.

A naive approach to resolving the problem of intersecting Cliques is to execute Cliques sequentially. For instance, in scenario (a) of Fig.~\ref{fig:intersecting_cliques}, if the top-left Clique runs first and corrects the length-1 error indicated by parity qubits `a' and `p', the syndrome is then updated such that only parity qubit `r' remains activated. The Clique decoder subsequently identifies this remaining syndrome (a single parity qubit is set on the surface code) as complex, which is the correct identification. However, this sequential approach is inefficient. Specifically, for a surface code with a code distance of $d$, it would require sequential execution of $d^{2}$ Cliques.

\begin{figure}[t]
\centering
\includegraphics[width=0.95\columnwidth,trim={0cm 0cm 0cm 0cm},clip]{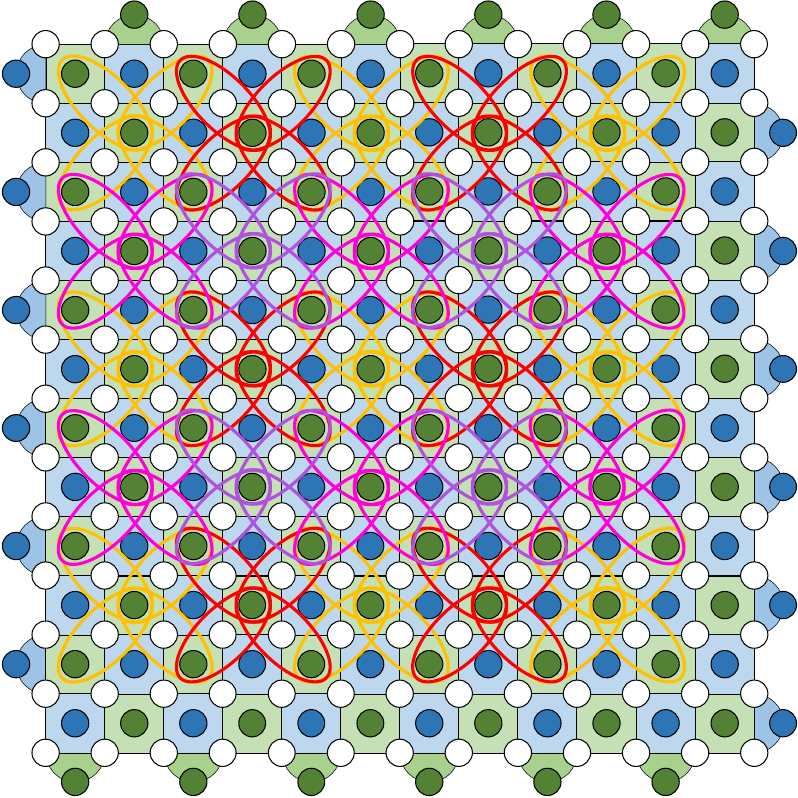}
\caption{Example illustrating parallel execution of non-intersecting Cliques (coverage of edges and corners is not shown): A-Yellow, B-Red, C-Pink. D-Purple; Cliques of the same color can be executed in parallel. Thus, a 4-step sequential process is sufficient to cover the entire surface code lattice.}
\label{fig:parallel_cliques}
\end{figure}

However, this potential imitation can be easily overcome by a parallelization approach.
The key idea is that the sequential execution described above is only required for intersecting Cliques.
It is intuitive that for the surface code, for any chosen Clique there are up to eight intersecting neighbors Cliques (i.e. the N, W, S, E, NW, SW, SE, NE neighbors).
This can be modeled as a simple graph coloring problem on an 8-connected grid which can be solved with 4 colors (rows alternating between $A-B-A-B$ and $C-D-C-D$ colors).
In other words, a 4-step sequential approach is sufficient to execute all the intersecting Cliques independently irrespective of the code distance (as shown in Fig.~\ref{fig:parallel_cliques}). In this 4-step approach, all the Cliques with the same color can be executed in parallel.


\subsection{Algorithmic Advantages of \texttt{Clique\_L2}}

\begin{figure}[t]
\centering
\includegraphics[width=0.95\columnwidth,trim={0cm 0cm 0cm 0cm},clip]{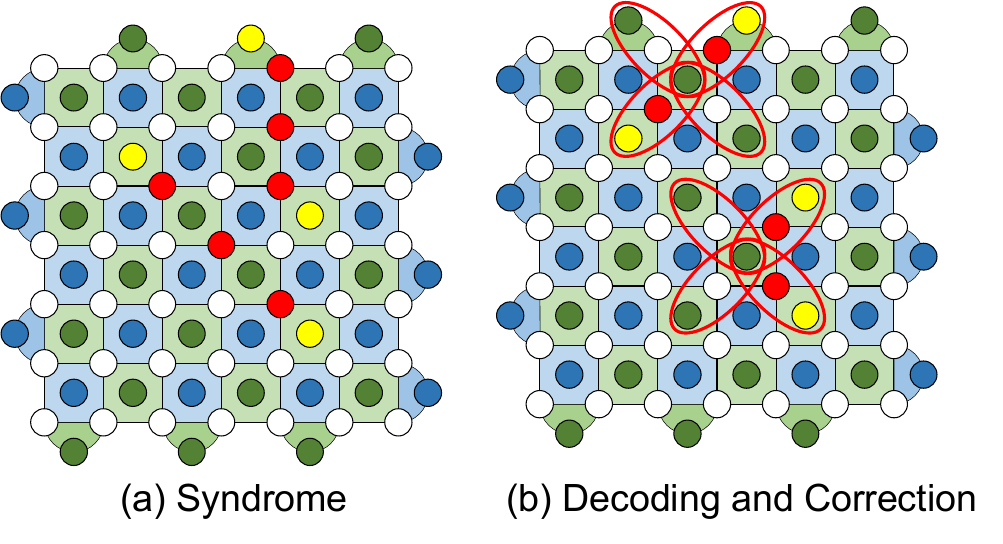}
\caption{An error configuration example. (a) The initial error configuration has 6 erring data qubits and it causes syndromes labeled in yellow. (b) The enhanced Clique decoder separates the syndromes into two groups and corrects them as error chains of length 2. 
}
\label{fig:decoding_shortest}
\end{figure}

The correction produced by the Clique-based approach aligns with the output of full-fledged decoders that aim to identify the most probable error configurations. For example, the error configuration in Fig.~\ref{fig:decoding_shortest}(a) consists of two error chains, each of length 3, resulting in the activation of the parity qubits shown in yellow. In the \texttt{Clique\_L2} decoding process, the syndromes are divided into two groups and addressed by two separate Cliques. The decoder then applies corrections to the data qubits marked in red, as illustrated in Fig.~\ref{fig:decoding_shortest}(b). As discussed in Section~\ref{BTWC_motive}, the error configuration shown in Fig.~\ref{fig:decoding_shortest}(b) is $10^{2}$ more likely than that in Fig.~\ref{fig:decoding_shortest}(a), since it involves two fewer erroneous data qubits overall.

\subsection{Measurement Errors}
The strategy for handling measurement errors is consistent with that of \texttt{Clique\_L1}. In our approach, measurement errors induce stochastic flips in the parity check qubits. By collecting syndrome data over multiple rounds, persistent errors will be attributed to data qubit errors. In this study, we process data from two measurement rounds, though additional rounds can be incorporated using the same methodology.

\section{Methodology}
\label{sec:method}


We compare the performance of \texttt{Clique\_L1} and \texttt{Clique\_L2} by measuring the percentage of syndrome data forwarded to full-fledged decoders over $10^8$ randomly sampled execution cycles per logical qubit.
X-type and Z-type errors are corrected independently, so focusing on either one is sufficient for modeling purposes. 
The local decoder analyzes the error signatures to determine whether the syndromes should be processed locally or need to be sent to a full-fledged decoder.


Based on the simulation scheme described, three noise models—the uniformly random, Gaussian, and Dual-Error models are applied. The latter two are implemented to simulate correlated noise in the form of clustered qubit errors. These models are applied before generating the parity check qubit syndromes. 

The uniformly random model is a typical phenomenological noise model for data qubit errors and measurement errors~\cite{AFS}.
Random errors are introduced on the data qubits of a surface code patch, following a specified physical error rate. Then random measurement errors are applied to the parity check qubits at the same rate.

\begin{figure}[t]
\centering
\includegraphics[width=0.95\columnwidth,trim={0cm 0cm 0cm 0cm},clip]{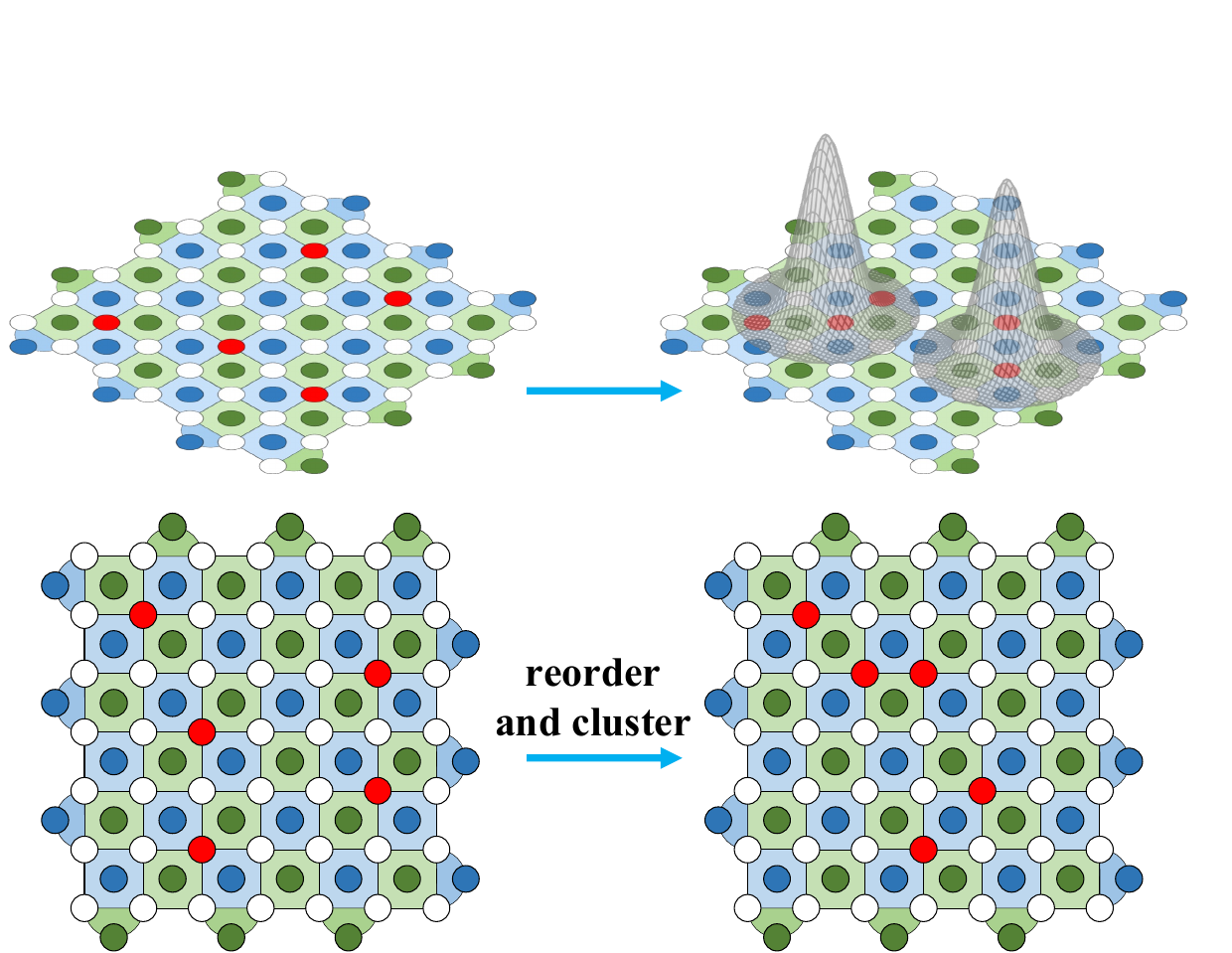}
\caption{Design of Gaussian model.}
\label{fig:gaussian_model}
\end{figure}


A Gaussian noise model describes a scenario in which qubit errors tend to cluster in groups and additional errors are more likely to occur near an existing error. 
Such clustered errors can be caused by fabrication defects, mis-directed qubit frequencies, two-level systems (TLS), burst errors caused by cosmic rays, and more~\cite{fuhui2023codesign, chadwick2024averting, leroux2024snakes}. 
To simulate noise at a given physical error rate, we first generate an error signature based on that rate, then group the erroneous qubits into clusters, as shown in Fig.~\ref{fig:gaussian_model}. We assume that the probability of errors occurring within each cluster follows a Gaussian distribution, meaning that qubits closer to the cluster center are more likely to be erroneous.

\begin{figure}[t]
\centering
\includegraphics[width=0.95\columnwidth,trim={0cm 0cm 0cm 0cm},clip]{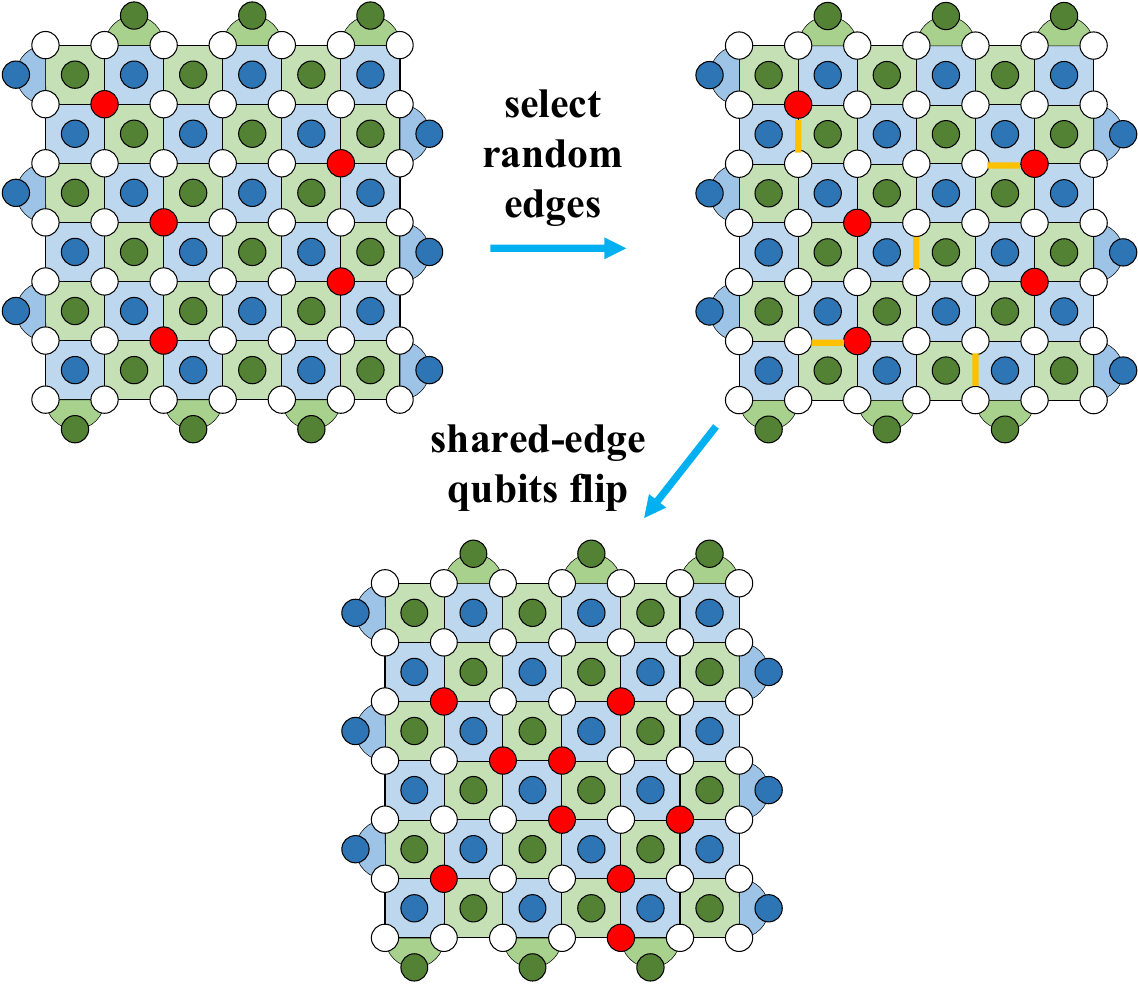}
\caption{Design of Dual-Error model.}
\label{fig:hook_model}
\end{figure}


The Dual-Error model is used to simulate scenarios where two adjacent data qubits are flipped. These could be a result of error clustering as described above as well as more specific coupling errors caused by error propagation through the CNOT gates in the circuits used to implement the syndrome checks. While there are specific nuances to how these errors propagate and how the syndrome events manifest, we model the errors more generally.
For the model, as illustrated in Fig.~\ref{fig:hook_model}, once an error signature is generated, edges on the surface code patch are randomly selected according to the physical qubit error rate. The data qubits connected by these selected edges are then flipped.

\section{Evaluation}

To evaluate the transmission bandwidth of \texttt{Clique\_L1} and \texttt{Clique\_L2}, we measure the fraction of decodes that are sent to full-fledged decoders in the presence of different levels of noises. We compare the performances of \texttt{Clique\_L1} and \texttt{Clique\_L2} for different physical error rates across a wide range of code distances. Then starting from code distance 19, we apply the logistic model to extrapolate the data up to code distance 25 since the trends persist.

\subsection{No Measurement Error}

\begin{figure}[t]
\centering
\includegraphics[width=0.95\columnwidth,trim={0cm 0cm 0cm 0cm},clip]{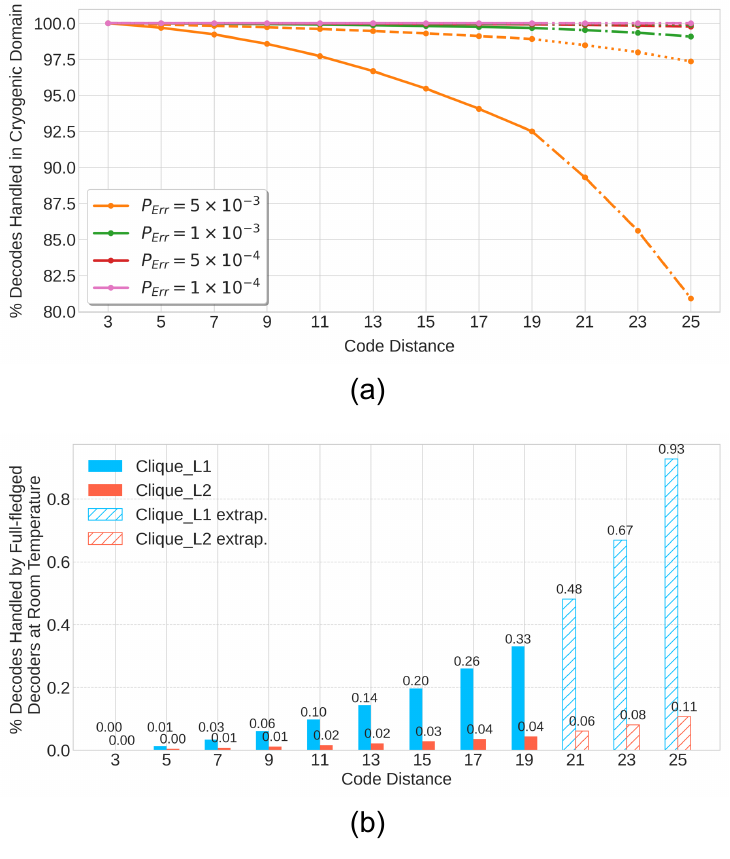}
\caption{Decodes handled by Cliques under no measurement error: (a) Fraction of decodes handled locally within cryogenic domain by Cliques for different physical error rates. Solid line (code distance 3–19): \texttt{Clique\_L1} simulation data; dashed line (code distance 3–19): \texttt{Clique\_L2} simulation data; dash-dot line (code distance 21–25): \texttt{Clique\_L1} extrapolated data; dotted line (code distance 21–25): \texttt{Clique\_L2} extrapolated data. (b) Fraction of decodes sent to full-fledged decoders at room temperature by Cliques at the physical error rate 0.1\%. }
\label{fig:no_measurement_original}
\end{figure}

When noise arises solely from data qubit errors, \texttt{Clique\_L2} significantly reduces the fraction of decodes that must be sent to room-temperature decoders, particularly at higher code distances. For example, at code distance 21—where each logical qubit encounters approximately 200 X or Z data error sources—and a physical error rate of 0.5\%, \texttt{Clique\_L2} offloads only about 1.52\% of decodes, compared to 10.70\% for \texttt{Clique\_L1}, yielding a 7.03x improvement (see Fig.~\ref{fig:no_measurement_original}(a)). This performance advantage persists across a range of physical error rates, with benefits ranging from 3.10x to 8.95x, and is especially pronounced at larger code distances. Notably, in Fig.~\ref{fig:no_measurement_original}(b), at a 0.1\% error rate and code distance 25, \texttt{Clique\_L1} offloads 0.93\% of decodes, whereas \texttt{Clique\_L2} reduces this to just 0.11\%, resulting in an 8.68x improvement. While complex error chains are less common at lower physical error rates, length-2 error chains remain likely at larger code distances—where \texttt{Clique\_L2} provides more effective mitigation of bandwidth demands.

\subsection{Uniformly Random Noise Model}

\begin{figure}[t]
\centering
\includegraphics[width=0.95\columnwidth,trim={0cm 0cm 0cm 0cm},clip]{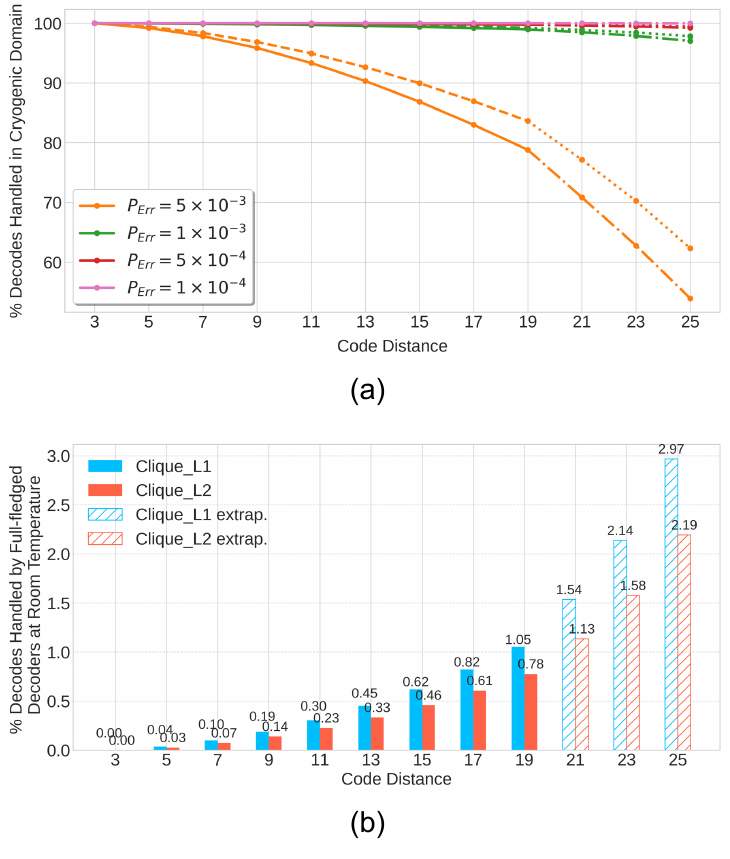}
\caption{Decodes handled by Cliques under uniformly random noise model: (a) Fraction of decodes handled locally within cryogenic domain by Cliques for different physical error rates. Solid line (code distance 3–19): \texttt{Clique\_L1} simulation data; dashed line (code distance 3–19): \texttt{Clique\_L2} simulation data; dash-dot line (code distance 21–25): \texttt{Clique\_L1} extrapolated data; dotted line (code distance 21–25): \texttt{Clique\_L2} extrapolated data. (b) Fraction of decodes sent to full-fledged decoders at room temperature by Cliques at the physical error rate 0.1\%.}
\label{fig:two_measurement_original}
\end{figure}

In the presence of uniformly random noise, \texttt{Clique\_L2} continues to reduce the fraction of decodes that must be sent off the cryogenic domain, with the benefits becoming more significant at larger code distances. For example, in Fig.~\ref{fig:two_measurement_original}(b), at a 0.1\% error rate and code distance 25, \texttt{Clique\_L1} sends approximately 2.97\% of decodes off the cryogenic domain, whereas \texttt{Clique\_L2} sends about 2.19\%, resulting in a 1.36x improvement. Across all code distances and physical error rates, \texttt{Clique\_L2} consistently handles a larger portion of errors locally, with benefits in the ratio of 1.22x to 1.44x. It is worth noting that the current measurement error mitigation logic may diminish the occurrence of length-2 errors, and further improvements in error handling could unlock even greater potential for \texttt{Clique\_L2}.

\subsection{Gaussian Noise Model with Measurement Error}

\begin{figure}[t]
\centering
\includegraphics[width=0.95\columnwidth,trim={0cm 0cm 0cm 0cm},clip]{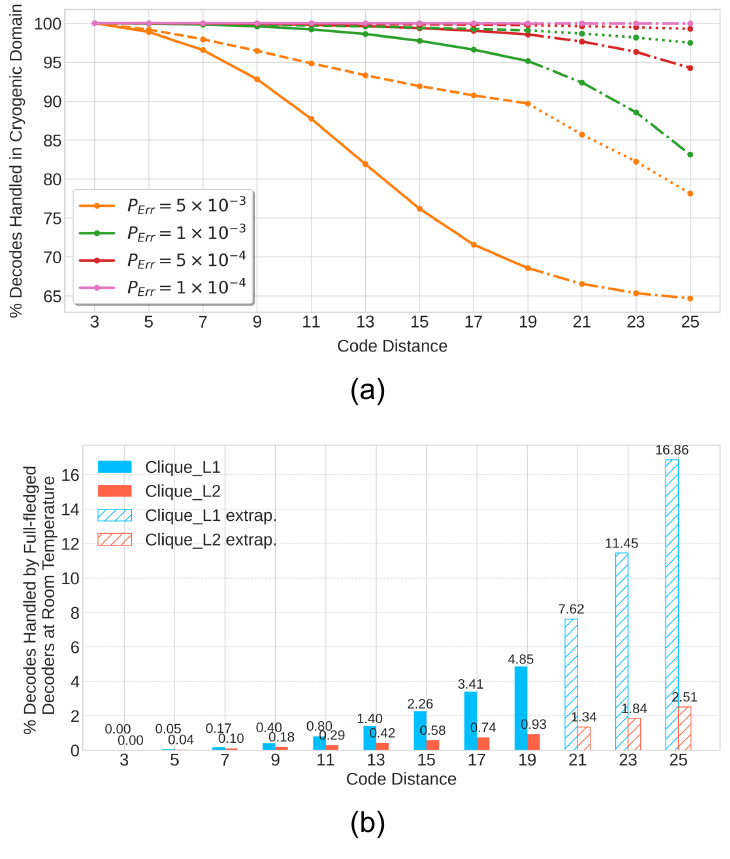}
\caption{Decodes handled by Cliques under measurement error and Gaussian noise Model: (a) Fraction of decodes handled locally within cryogenic domain by Cliques for different physical error rates. Solid line (code distance 3–13): \texttt{Clique\_L1} simulation data; dashed line (code distance 3–13): \texttt{Clique\_L2} simulation data; dash-dot line (code distance 15–31): \texttt{Clique\_L1} extrapolated data; dotted line (code distance 15–31): \texttt{Clique\_L2} extrapolated data. (b) Fraction of decodes sent to full-fledged decoders at room temperature by Cliques at the physical error rate 0.1\%.}
\label{fig:two_measurement_Gaussian}
\end{figure}

In the Gaussian noise model, errors tend to cluster more densely compared to the uniformly random model. As a result, \texttt{Clique\_L2} achieves a more substantial reduction in data transmission bandwidth between the cryogenic environment and room temperature, with improvements ranging from 1.36x to 8.60x—particularly at larger code distances (see Fig.~\ref{fig:two_measurement_Gaussian}(a)). At code distance 25, each logical qubit experiences approximately 1300 error sources, stemming from both data qubits and measurement operations. In Fig.~\ref{fig:two_measurement_Gaussian}(b), at a physical error rate of 0.1\%, the likelihood of clustered and complex error chains increases. In this setting, \texttt{Clique\_L1} offloads nearly 16.86\% of decodes, whereas \texttt{Clique\_L2} processes most cases locally, offloading only about 2.51\%, resulting in a 6.71x improvement.

\subsection{Dual-Error Noise Model with Measurement Error}

\begin{figure}[t]
\centering
\includegraphics[width=0.95\columnwidth,trim={0cm 0cm 0cm 0cm},clip]{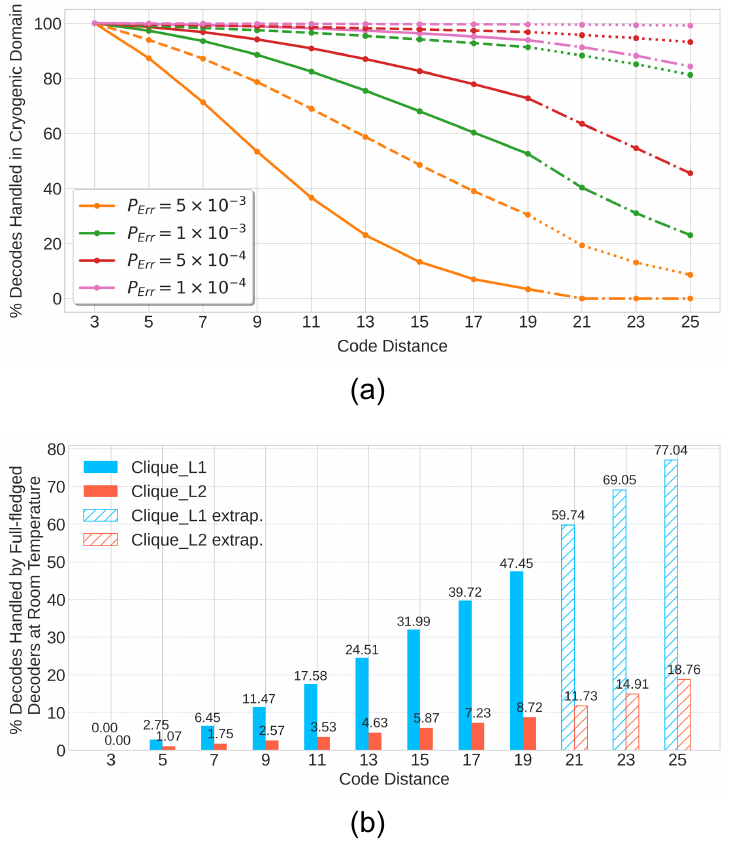}
\caption{Decodes handled by Cliques under measurement error and Dual-Error noise Model: (a) Fraction of decodes handled locally within cryogenic domain by Cliques for different physical error rates. Solid line (code distance 3–13): \texttt{Clique\_L1} simulation data; dashed line (code distance 3–13): \texttt{Clique\_L2} simulation data; dash-dot line (code distance 15–31): \texttt{Clique\_L1} extrapolated data; dotted line (code distance 15–31): \texttt{Clique\_L2} extrapolated data. (b) Fraction of decodes sent to full-fledged decoders at room temperature by Cliques at the physical error rate 0.1\%.}
\label{fig:two_measurement_Hook}
\end{figure}

Under the Dual-Error noise model, long error chains and large clusters of errors occur more frequently than in the Gaussian model. As a result, \texttt{Clique\_L1} faces substantial challenges: at a physical error rate of 0.5\%, nearly all decoding tasks must be forwarded to full-fledged decoders starting from code distance 15. In contrast, \texttt{Clique\_L2} continues to handle most decoding tasks locally even at code distance 25, except at the highest error rate (0.5\%). Across nearly all error rates, \texttt{Clique\_L1} consistently offloads the majority of decoding tasks to room-temperature decoders. As shown in Fig.~\ref{fig:two_measurement_Hook}(a), the improvement offered by \texttt{Clique\_L2} ranges from 1.38x to 18.38x across all code distances and error rates. Specifically, Fig.~\ref{fig:two_measurement_Hook}(b) highlights that at a 0.1\% error rate, \texttt{Clique\_L2} reduces bandwidth usage by 2.58x to 5.49x compared to \texttt{Clique\_L1} across all code distances.

\section{Discussion}

\texttt{Clique\_L2} achieves extended functionality which includes length-2 errors without requiring significant modifications. This enhanced capability proves effective at larger code distances. Moreover, in scenarios where complex error patterns, such as clustered errors, are more likely to occur, \texttt{Clique\_L2} demonstrates robust performance by handling a greater proportion of decoding locally and reducing transmission bandwidth across a wide range of physical error rates. Given that \texttt{Clique\_L1} was already considered promising for near-term QEC devices like superconducting transmon quantum systems, the extended design of \texttt{Clique\_L2} holds significant potential for practical applications.

\texttt{Clique\_L2} incurs some additional hardware overhead. The added logic is minimal, just a few extra gates per decoding unit. The main cost arises from pipelining required to support the four-stage decoding logic, as well as the 4-step ordering of Clique decodings across the logical qubit patch. However, this cost remains modest due to the small data width involved. Overall, the substantial bandwidth reduction achieved by \texttt{Clique\_L2} far outweighs its hardware cost.

\subsection{Future Work}

\circled{1}\ The measurement error mitigation algorithm could be enhanced to recover more complex error patterns. Currently, after two rounds of measurement, persistent errors in parity qubits are attributed to data qubit errors. However, this approach may overlook complex error patterns, as not all errors remain clustered after multiple measurement rounds.

\circled{2}\ The algorithm of the \texttt{Clique\_L2} is designed to be implementable within the cryogenic domain, and thus the corresponding hardware logic remains to be investigated. 

\circled{3}\ Exploring localized decoding for length‑$k$ errors using the Cliques framework is a promising approach. In this context, Cliques would be applied to manage error chains with length $k>2$. This involves extending the leaves of the Cliques to cover a larger number of qubits, while simultaneously adapting the decoding logic. The trade-offs between increased logic complexity and data transmission bandwidth remain to be thoroughly investigated. However, this localized decoding approach using Cliques could also offer valuable insights for full-fledged decoding algorithms.

\subsection{Related Work}

Numerous room-temperature second-level decoders have been proposed \cite{delfosse2021almost, liyanage2023scalable, barber2025real, wu2023fusion, vittal2023astrea, AFS, das2022lilliput}. Our approach is orthogonal to these designs and can be adapted to work with any of them.
Although hardware design is not the focus of this work, ongoing advancements in cryo-CMOS \cite{bardin201929, chakraborty2022cryo, underwood2024using, van2020scalable, park2021fully, kossel202440, hinderling2024flip, bersano2023quantum, frank2023low, bardin2019design, zou2021frequency}, SFQ technologies \cite{kim2024fault, choi2024supercore, holmes2020nisq+, jokar2022digiq, ishida2020supernpu, kashima202164}, and modeling tools \cite{byun2022xqsim, min2022cryowire, min2023qisim} can be leveraged to support and extend our framework.
Several local decoder designs have been proposed \cite{chamberland2023techniques, delfosse2020hierarchical, alavisamani2024promatch, ravi2023better, holmes2020nisq+, smith2023local}, but most are not sufficiently lightweight for cryogenic deployment.

\section{Conclusion}

In conclusion, this work demonstrates a significant improvement in Clique decoding by introducing \texttt{Clique\_L2}, an enhanced local decoder capable of correcting both trivial (length-1) and simple correlated (length-2) error chains in the cryogenic domain. Compared to \texttt{Clique\_L1}, \texttt{Clique\_L2} substantially reduces the I/O bandwidth required for decoding across multiple realistic noise scenarios. Compared to \texttt{Clique\_L1}, \texttt{Clique\_L2} improves bandwidth demands by up to 8.95x for data-qubit-only errors, 1.44x under uniformly random noise, and 8.60x under densely clustered Gaussian error conditions. Moreover, in the challenging Dual-Error noise scenario characterized by frequent long and complex error chains, \texttt{Clique\_L2} demonstrates exceptional efficiency by locally resolving up to 18.38x decoding tasks compared to \texttt{Clique\_L1}. 

These improvements highlight the feasibility and effectiveness of incorporating slightly more sophisticated cryogenic hardware to significantly mitigate the classical decoding bottleneck, contributing to progress toward more scalable and robust fault-tolerant quantum computation.

\section*{Acknowledgment}

This material is based upon work supported by the U.S. Department of Energy, Office of Science, Office of Advanced Scientific Computing Research, Accelerated Research in Quantum Computing under Award Number DE-SC0025633. This research used resources of the National Energy Research Scientific Computing Center, a DOE Office of Science User Facility supported by the Office of Science of the U.S. Department of Energy under Contract No. DE-AC02-05CH11231 using NERSC award ASCR-ERCAP0033197.

\bibliographystyle{IEEEtranS}
\bibliography{references,ref2}

\end{document}